\documentstyle[11pt, psfig]{article}
\begin{document}
\begin{center}
{\Large\bf Using Conditional Phase Change Operations to Probe the Nature of 
Quantum Mechanical Systems}
\vspace{0.2in}

\centerline{Sarnath 
Ramnath$^{\dag\,}$\footnote{Email: sarnath@eeyore.stcloudstate.edu} and 
Kevin Haglin$^{\ddag\,}$\footnote{Email: haglin@stcloudstate.edu}}
\centerline{\it $^{\dag}$Department of Computer Science, St. Cloud State 
University}
\centerline{\it 720 Fourth Avenue South, St. Cloud \ MN 56301, USA}
\vskip 1 \baselineskip
\centerline{\it $^{\ddag}$Department of Physics and Astronomy, St. Cloud 
State University}
\centerline{\it 720 Fourth Avenue South, St. Cloud \ MN 56301, USA}
\end{center}

\vskip 1 \baselineskip
\centerline{\bf Abstract}
Recent developments in quantum computing suggest that it could
be possible to make conditional changes to the state of a quantum
mechanical system without resorting to classical observation.  It
is accomplished through collective response of atoms comprising a 
lattice for the system and involves relative phase adjustments only.
We exploit this possibility and describe experimental designs that 
could help elucidate quantum mechanical properties of these
systems and distinguish between interpretations. 

\section{Introduction}

A long-standing question in quantum mechanics is the following:
If a measurement shows a quantum system to be in a state $s$, in what state 
was the system before the measurement was made? There are three
widely recognized plausible answers to this question that characterize the 
main schools of thought:
\begin{enumerate}
\item{} The system was in the state $s$, but we were unable to determine that 
due to insufficient information.
\item{} The system was not in any particular state, and it was the act of 
measurement that forced it to choose a state in a quasi-random 
manner, respecting the probability distribution. 
\item{} The question is irrelevant since we have no way of verifying what 
state the system was in prior to making the measurement.

\end{enumerate}

Until now, it has been widely believed that this question cannot be resolved 
by experiment. However, recent developments in the area of quantum computing 
have shown that it is possible to perform certain transform operations 
on quantum systems without observing the state---without disturbing the
wave function. In this article we 
describe a possible experiment using these operations that may finally 
enable us to elucidate this question. In addition to the philosophical 
issues, this experiment may actually help us design other similar 
ones that can be used to decipher the mysteries of quantum mechanics.

\section{Technique employed}

The operation that we propose to use is the so called ``Conditional Phase 
Flip" (CPF), described below.

Given: A quantum system in some superposition of $n$ states 
$s_1, s_2, ..., s_n$, and a (possibly) unknown but fixed value $e$, which 
equals some $s_i$, $1 \leq i \leq n$.

Output: A quantum system that is in the same superposition as the 
input, with the exception that the phase of $s_i$ is rotated by $\pi$ 
(i.e., inverted, in some sense).

This operation is a specific instance of the more general technique of
Conditional Phase Shift.  The ability to accomplish CPFs has been 
demonstrated using amplitude resonance of atomic lattice
center-of-mass motion probed phononically\cite{iontrap}. 
Better, in terms of timescales, would be to accomplish the feat
fully optically.  Progress in this direction has also been 
documented\cite{qat95}.
The remarkable feature of these types of manipulation is that they
seem to be able to carry out a CPF without 
resorting to a classical observation.
Our purpose here is not to review the technology of
cooling and trapping ions for construction of quantum gates\cite{bek98}, but 
instead we assume the existence and reliability of such tools 
for use in experiments that allow
fundamental studies of the nature of quantum mechanics.

Models employing this physical operation have been used to design quantum 
algorithms, most notably the quantum database searching algorithm of 
Grover \cite{grover}. Grover combines the CPF with an ``inversion about mean''
operation, which can be achieved using the Walsh-Hadamard transform 
to increase the probability for the system to be found in a desired state, 
starting with a uniform distribution.
Inversion about the mean is best understood as follows. Consider a 
quantum system in a superposition of $n$ states, $s_1, s_2, ... s_n$,
with amplitudes $a_1, a_2, ..., a_n$, respectively. Further, for any 
two states $s_i$ and $s_j$, $1 \leq i, j \leq n$, $s_i$  and $s_j$ are 
either in phase or have a phase difference of $\pi $. The phase thus 
induces a partitioning of the states into two subsets. Without loss of 
consistency, we may assign a phase of $0$ to one set and a phase of 
$\pi $ to the other. Since $e^{i0} = 1$ and $e^{i\pi} = -1$, we can think 
of each $s_i$ as having 
a ``signed'' amplitude, $+a_i$ if the phase is $0$ and $-a_i$ if the 
phase is $\pi $. These signed amplitudes can be represented by means of 
a histogram. Let $\mu $ denote the arithmetic mean of the signed amplitudes
of all the $n$ states. An inversion about the mean operation causes the 
histogram to be reflected about the line $y = \mu $.
To date, there does not appear to be any 
successful attempt implementing this algorithm in all its complexity, 
though other quantum algorithms (viz., Shor's quantum factoring algorithm)
have been tested using ion traps \cite{iontrap}.

\section{Designing the set up}

In our proposed experiment, we use a simple quantum system consisting of 
2 qubits. In the article by Cleve et. al. \cite{revisited}
a simple generalized framework for representing quantum algorithms 
is discussed. The two key components in this framework are the 
Half-Silvered Mirror (HSM) and the Conditional Phase Shifter (CPS).
Suppose we wish to force our 2-qubit system into a desired state, 
say $| 11\rangle $. Initially the system is in 
state $| 00\rangle $, i.e., we have 2 qubits, each in the state $| 0\rangle $.

\begin{figure}
\centerline{\psfig{file=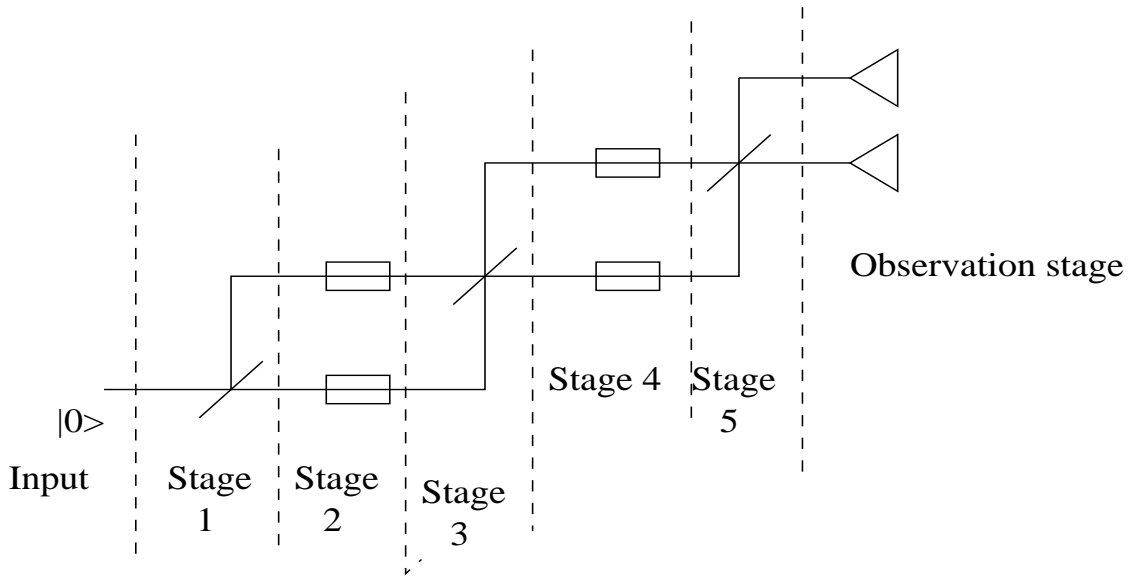,height=3in,width=6in}}
\caption{ Quantum System that converts $| 00\rangle $ to $| 11 \rangle$}
\label{f1}
\end{figure}

Figure \ref{f1} shows how this set up will look for each of the 
two qubits.
In the first stage we employ a HSM (one for each qubit) to 
obtain the superposition $\frac{1}{2}(| 00\rangle + |01\rangle +  |10\rangle + |11\rangle)$. 
Note that each HSM performs the transform 
\begin{eqnarray}
|0\rangle & \rightarrow &  \frac{1}{\sqrt{2}}\left(|0\rangle + |1\rangle\right)
\nonumber\\
|1\rangle &\rightarrow & \frac{1}{\sqrt{2}}\left(|0\rangle - |1\rangle\right).
\end{eqnarray}
Since we are in the initial state $|00\rangle$, we have the required result.
The second stage employs a CPS that shifts the phase of $|11\rangle$ (which is 
our desired state) by $\pi $, to yield the superposition
$\frac{1}{2}(|00\rangle + |01\rangle +  |10\rangle - |11\rangle)$. 

The last three stages employ an HSM, a CFS and an HSM, respectively. 
The CFS in stage 4 shifts the phase of $|00\rangle$. These three steps 
are considered as a group, since their combined effect is to 
cause an inversion about the mean \cite{grover}. The mean of the 
amplitudes of the four states at the end of stage 2 is 1/4. 
This means that the signed amplitude values of each state are reflected 
about the line $ y = \frac{1}{4}$. A point 
$(x, \frac{1}{2})$ is reflected as $(x, 0)$  and the point
$(x, \frac{-1}{2})$ is reflected as $(x, 1)$.Therefore, the inversion 
about the mean causes the amplitudes of $|00\rangle$, $|01\rangle$ and 
$|10\rangle$ to drop to 
zero, leaving the system in the state  $|11\rangle$.  This process, as 
explained in \cite{revisited}, is achieved by 2 crossed paths for 
each qubit, one path
representing a 0 and the other path representing a 1. The expected behavior 
of this system is as predicted by QED.

\begin{figure}
\centerline{\psfig{file=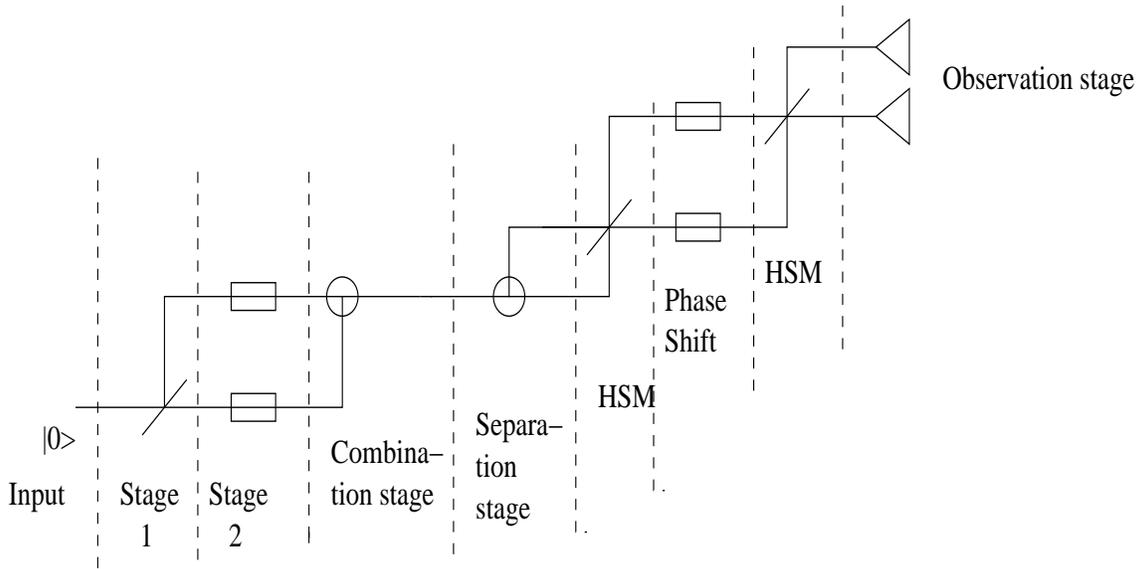,height=3in,width=6in}}
\caption{ Proposed Quantum System for our experiment}
\label{f2}
\end{figure}

What we have described above is essentially a simple version of Grover's 
quantum search algorithm, where we are ``searching'' for the  pre-defined 
state $|11\rangle$. For our experiment, we modify the the set up as shown in 
figure \ref{f2}.
In the new set up, 
after stage 2, we simply combine the two paths representing the 
state of each qubit into one, i.e., a simple additive 
superposition, instead of using the interference caused by the 
HSM. The result of this is a system in the exact same superposition, 
i.e., $\frac{1}{2}(|00\rangle + |01\rangle +  |10\rangle - |11\rangle)$, as 
we had earlier, except that each qubit is now traveling on a single path.
We then employ the last three stages as described previously. 

\section{Expected results}

The last three stages will cause an inversion about mean.
Consider the two possible scenarios:

1. If the system were in any one of the four possible states, then the 
final state of the system will be of the form 
$\frac{1}{2}(\pm |00\rangle \pm |01\rangle \pm  |10\rangle \pm |11\rangle), $
depending on the initial state. 
For instance, if the system were in state $| 00 \rangle$,  the mean is
$\frac{1}{4}$ and inversion about the mean implies reflection about the 
line $ y = \frac{1}{4}$. This would assign the state $| 00 \rangle$ an 
amplitude of $\frac{-1}{2}$  and the other states an amplitude of 
$\frac{1}{2}$. (The three other states 
will always end up with a phase opposite to that of the initial state.)
In other words, the system at the end of the experiment, is equally likely 
to be in any one of the four states. 

2. If, on the other hand, the system were in a superposition of the 
states (and that it is the act of measurement that forces the system into a 
state), the final state of the system has to be $|11\rangle$, as explained in 
the previous section.

\section{Another possible approach}

\begin{figure}
\centerline{\psfig{file=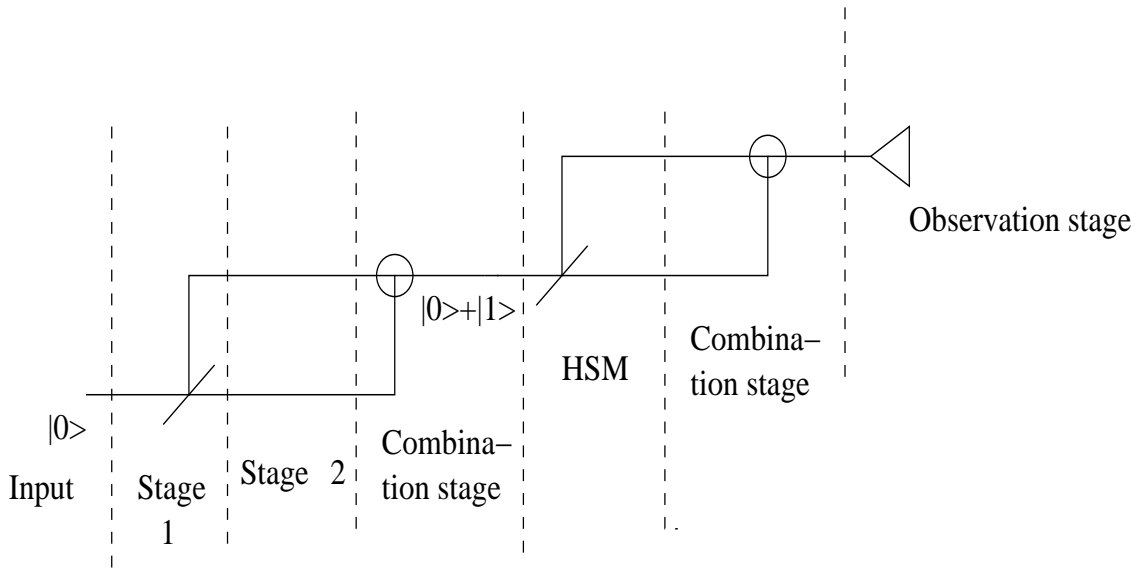,height=3in,width=6in}}
\caption{Alternate set up using one qubit}
\label{f3}
\end{figure}

Another possible approach would be a set up described in Figure \ref{f3}.
In this case we employ only one qubit and do not have a separation stage. 
The first three stages are identical to what we had earlier, with the 
exception that there are no phase shifts employed on any of the paths. 
In the 
fourth stage, instead of separating the states, we allow the photon 
in the superposition ${1\over\sqrt{2}}\left(|0\rangle + |1 \rangle\right)$ 
to fall on the HSM.  If the Copenhagen 
interpretation were to be followed, the $|0\rangle$ state results 
in a $|0\rangle$ state for the lower path with amplitude
1/2 and a $|1\rangle$ state 
for the upper path with amplitude 1/2, whereas the $|1\rangle$ state 
results in a 
$|0\rangle$ for the upper path and a $-|1\rangle$ state for the 
lower path, again with amplitudes of 1/2. This leaves the lower 
path in the superposition 
${1\over\/2}\left(|0\rangle - |1\rangle\right)$, 
and the upper path in the superposition
${1\over\/2}\left(|0\rangle + |1\rangle\right)$. In the fifth stage
we combine the paths once again in an additive manner as explained 
earlier. If the phase could be controlled, this would leave the 
system in the final state $|0\rangle$. 

If, on the other hand, we assume that in stage three the system were either in
state $|0\rangle$ or in state $|1\rangle$ (but not both simultaneously), 
this would imply that after 
stage four, the system is either in state ${1\over\sqrt{2}}\left(|0\rangle
+ |1\rangle\right)$ or in state ${1\over\sqrt{2}}\left(|0\rangle - 
|1\rangle\right)$, respectively. In either case, 
a measurement would be equally likely to find the system in state 
$|0\rangle$ or in state $|1\rangle$. Such a measurement can be made
after the fifth stage using a polaroid filter.
 
\section{Discussion}

The techniques described above enable us to affect 
quantum systems, which have different expected results, depending upon the
nature of the systems' state vector prior to measurement. Measuring the 
final state enables us to say 
something about the state in which the system existed prior to 
observation. 
If such experiments are shown to be feasible in practice, 
it would no longer be appropriate to retreat to an ``agnostic'' position, 
since we would have ways of knowing the state of a system prior to
its direct measurement.  Furthermore, these types of experiments
have another potential benefit in terms of tapping additional
power within quantum computers.
It may be possible to generalize these 
ideas to develop other experiments that would get us closer to 
understanding the nature of quantum mechanical systems.

\end{document}